\documentstyle[12pt]{article}
\def\be{\begin{equation}}
\def\ee{\end{equation}}
\def\ben{\begin{displaymath}}
\def\een{\end{displaymath}}
\def\ba{\begin{array}{c}}
\def\ea{\end{array}}
\def\bea{\begin{eqnarray}}
\def\eea{\end{eqnarray}}

\begin{document}

\sloppy

\titlepage

\begin{center}{\Large \bf

${\mathcal CPT}-$conserving Hamiltonians and their

nonlinear supersymmetrization

using differential charge-operators ${\mathcal C}$

 } \end{center}

\begin{center} {\bf
 Bijan Bagchi} and {\bf A. Banerjee}
% \vspace{3mm}

Department of Applied Mathematics,
 University of Calcutta

92 Acharya Prafulla Chandra Road, Kolkata 700 009,
 West Bengal, India

e-mail: bbagchi123@rediffmail.com

 \vspace{3mm} {\bf
 Emanuela Caliceti}

% \vspace{3mm}

Dipartimento di Matematica dell' Universit\`{a} and Istituto
Nazionale di Fisica Nucleare, I-40127 Bologna, Italy

e-mail: caliceti@dm.unibo.it

 \vspace{3mm} {\bf Francesco Cannata}

% \vspace{3mm}

Dipartimento di Fisica dell'Universit\`a and
         Istituto Nazionale di Fisica Nucleare,
         I-40126 Bologna, Italy

e-mail: Francesco.Cannata@bo.infn.it

 \vspace{3mm} {\bf Hendrik B. Geyer}

% \vspace{3mm}

Institute of Theoretical Physics, University of Stellenbosch,

Matieland 7602, South Africa

e-mail: hbg@sun.ac.za

% \vspace{3mm}

 \vspace{3mm} {\bf
 Christiane Quesne}

% \vspace{3mm}

Physique Nucl\'eaire Th\'eorique et Physique Math\'ematique \\
Universit\'e Libre de Bruxelles, Campus de la Plaine CP229, B-1050
Brussels, Belgium

e-mail: cquesne@ulb.ac.be

 \vspace{3mm}

and
 \vspace{3mm}

{\bf Miloslav Znojil}

% \vspace{3mm}

\'{U}stav jadern\'e fyziky AV \v{C}R, 250 68 \v{R}e\v{z}, Czech
Republic

e-mail: znojil@ujf.cas.cz

% \vspace{3mm}

 \end{center}

%\vspace{5mm}

\newpage

\vspace{5mm}
\section*{Abstract}

A brief overview is given of recent developments and fresh ideas
at the intersection of ${\cal PT}-$ and/or ${\cal CPT}-$symmetric
quantum mechanics with supersymmetric  quantum mechanics (SUSY
QM). Within the framework of the resulting supersymmetric version
of ${\cal CPT}-$symmetric quantum mechanics we study the
consequences of the assumption that the ``charge" operator ${\cal
C}$ is represented in a differential-operator form of the second
or higher order. Besides the freedom allowed by the Hermiticity
constraint for the operator  ${\cal CP}$, encouraging results are
obtained in the second-order case. In particular, the
integrability of intertwining relations proves to match the
closure of our nonlinear ({\it viz.,} polynomial) SUSY algebra. In
a particular illustration, our form of ${\cal CPT}-$symmetric SUSY
QM leads to a new class of non-Hermitian polynomial oscillators
with real spectrum which turn out to be ${\cal PT}-$asymmetric.

\vspace{5mm}

PACS 02.30.Tb  03.65Ca  03.65.Db  03.65.Ge

 \vspace{15mm}

 \subsection*{Keywords}

${\cal PT}-$symmetric Hamiltonians; ${\cal CPT}-$symmetric quantum
mechanics; supersymmetric  quantum mechanics; nonlinear SUSY
algebra; intertwining relations; ${\cal PT}-$asymmetric potentials

\newpage

\section{Introduction}

The recent growth of interest in the possibility of working with
non-Hermitian observables in quantum theory (cf. the concise
review papers \cite{BBRR})  is mainly due to the influential
Bender's and Boettcher's letter \cite{BB} where its authors
observed that the spectrum of certain Hamiltonians $H \neq
H^\dagger$ seems real and discrete and bounded below.

They conjectured that such an observation may find a firmer
mathematical background and explanation in the symmetry of their
models with respect to the combined action of the parity ${\cal
P}$ and time-reversal (i.e., complex conjugation) ${\cal T}$. This
inspiring idea has been further developed and re-formulated as
proposals of the so called ${\cal PT}-$symmetric quantum mechanics
\cite{BBjmp}, pseudo-Hermitian quantum mechanics \cite{205} and
${\cal CPT}-$symmetric quantum mechanics \cite{BBJ}. They all deal
with more or less the same class of the specific non-Hermitian
models characterized, in the language of the latter reference, by
another symmetry operator ${\cal C}$ which is very conveniently
called ''charge".

There exists an extensive literature on $\mathcal{PT}$-symmetric
quantum mechanics \cite{proceeda,proceed}. In particular, in a
number of papers \cite{Ca98,2341,23456,Ba02}, unexpected
consequences of the non-Hermiticity of Hamiltonians have been
noticed to emerge after its supersymmetrization a la Witten
\cite{Witten}. In terms of local models $H=p^2+V(x)$ on the real
line ($x \in I\!\!R$) where $V(x)=V^*(-x)$, these Hamiltonians
satisfy the intertwining relation
 \be
 H^\dagger\,{\cal P} = {\cal P}\,H\,.
 \label{PPH}
 \ee
Such $\mathcal{PT}$-symmetric Hamiltonians may have either
complex, or real spectra. When the $\mathcal{PT}$ symmetry remains
spontaneously unbroken and all the spectrum is real \cite{BB}, one
has elaborated the concept of quasi-Hermiticity of the Hamiltonian
\cite{Geyer,Kr04}. This means that the intertwining relation
(\ref{PPH}) holds also with $\mathcal{P}$ replaced by a
positive-definite operator $\Theta=\Theta^\dagger>0$ which plays
the role of a metric operator. The physical interpretation of such
models is standard \cite{Prugovecki}. When the spectrum is
complex, relation (\ref{PPH}) can still be written with
$\mathcal{P}$ replaced by a pseudo-metric
\cite{Dirac,pseudo,Ahmed}.

We shall now generalize the previous considerations to a new type
of symmetry. The framework of our constructions proposed in our
recent letter \cite{Ca04} will incorporate Hamiltonians with both
real and complex spectra. Correspondingly, we shall also deal, in
general, with non-positive metric (i.e., pseudo-metric). As for
the case of $\mathcal{PT}$ symmetry, the interpretation of this
type of quantum mechanics can be disputable \cite{Kr04} and might
require some innovation. However, we stress that, in our
framework, we can find models which have real spectra, where, in
particular, a non-Hermitian Hamiltonian is related by similarity
transformations not only to a Hermitian operator, but, more
specifically, to a Hermitian Schr\"odinger operator. Thus, for
these cases, we recover the standard quantum mechanics, after
similarity. Therefore, these cases are not disputable in their
interpretation. From a conservative point of view, one might
restrict the interest of our supersymmetric approach insofar as
one takes it instrumentally as a strategy to find complex
Hamiltonians with real spectrum that do not satisfy $\mathcal{PT}$
invariance (see Section \ref{Ema} below).

% \section{SUSY Quantum Mechanics revisited \label{SR}}

\subsection{SUSY intertwining relations}
\label{Inter}

In the same spirit as in Ref.~\cite{Ca04},  we shall study the
intertwining relations
 \be
 {\mathcal{F}}H^\dag = H{\mathcal{F}}\,
 \label{inter_rel}
 \ee
mediated by the Hermitian operator
 \be
\mathcal{F}=\mathcal{CP}\qquad (\mathcal{F}=\mathcal{F}^\dag)\,,
\label{F}
 \ee
where $\mathcal{P}$ is the parity operator, and $\mathcal{C}$ a
generalized ''charge" operator, assumed to be a polynomial in the
differential operator $d/dx$. For any Hamiltonian $H$, Eq.
(\ref{inter_rel}) is equivalent to $\mathcal{CPT}$ conservation,
with $\mathcal{T}$ the time reversal operator,
 \be
 {\mathcal{CPT}}H=H{\mathcal{CPT}}\,.
 \ee
In this paper we shall not discuss in detail the metric
interpretation for $\mathcal{F}$, but only stress the fact that,
if $\mathcal{F}$ and $H$ satisfy Eq. (\ref{inter_rel}), then also
${\mathcal{F}}^{-1}$ (if it exists) and $H$ meet an intertwining
 \be
 H^\dag{ \mathcal{F}}^{-1}= {\mathcal{F}}^{-1}H\,,
 \label{5}
 \ee
which means that $H$ is pseudo-Hermitian with respect to
${\mathcal{F}}^{-1}$, i.e.,
${\mathcal{F}}^{-1}-$pseudo-Hermitian~\cite{205}.

This observation may be useful for implementing the metric based
on $ {\mathcal{F}}^{-1}$, when $ {\mathcal{F}}^{-1}$ has a better
behavior than $\mathcal{F}$, {\it e.g.} with respect to
boundedness. Nevertheless, in our text we also use for a
$\mathcal{F}$ satisfying Eqs. (\ref{F}), (\ref{inter_rel}), the
word ``metric'' operator. In fact, Eq. (\ref{inter_rel}) implies
that $H\mathcal{F}$ is Hermitian. As a consequence of (\ref{5}),
if $\mid \phi \rangle$ and $\mid \psi \rangle $ are two arbitrary
vectors of the Hilbert space $L^2 (\mathbf R)$, we have
 \ben
 \int \phi^*(x)
 \left (
 {\mathcal{F}}^{-1}
 H\psi
\right )
 (x)dx =
 \int \psi^*(x)
 \left ( {\mathcal{F}}^{-1}
 H\phi
\right )
 (x)dx\,.
 \een
This can be interpreted as a Hermiticity condition for $H$
provided the scalar product is defined as
 \begin{eqnarray*}
\left\langle \phi \mid \psi \right\rangle_{{\mathcal{F}}^{-1}}
 & = & \int \phi^*(x)
 \left (
 {\mathcal{F}}^{-1}\psi
 \right )
 (x) \, dx \,;  \\
\left\langle \psi \mid \phi \right\rangle_{{\mathcal{F}}^{-1}} &
 = & \int \psi^*(x)
 \left (
 {\mathcal{F}}^{-1} \phi
 \right )(x)\, dx \,.
 \end{eqnarray*}
It is worthwhile to point out, however, that, in absence of
additional constraints, neither ${\mathcal{F}}^{-1}$ nor
$\mathcal{F}$ is necessarily positive definite so that, for
instance, the equation ${\mathcal{F}} \mid \phi \rangle = 0 $
might have a non-trivial solution different from $ \mid \phi
\rangle =0$. At this level, $ \langle \phi \mid \phi
\rangle_{{\mathcal{F}}^{-1}} $ does not define a true norm but
merely a pseudo-norm~\cite{pseudo,Ja02}.

It is evident that solving Eq. (\ref{inter_rel}) amounts to
analyzing the compatibility between $\mathcal{C}$ and $H$; in
other words, $\mathcal{C}$ and $H$ are to be found contextually.
Once Eq. (\ref{inter_rel}) is formally solved, one can investigate
its supersymmetrization~\cite{Sokolov}. By this we mean the
construction of super-charges
 \be
 Q=\left(  \begin{array}{cc} 0 & \mathcal{F} \\ 0 &  0
 \end{array} \right), \qquad
\tilde{Q}=\left(  \begin{array}{cc} 0 & 0 \\ {\mathcal{F}}^* & 0
 \end{array} \right)
\label{sup_Q}
 \ee
with anti-commutator
 \be K \equiv \left\{ Q, \tilde{Q} \right\}\,=\, \left(
\begin{array}{cc} {\mathcal{F F}}^* & 0 \\ 0 & {\mathcal{F}}^* \mathcal{F}
 \end{array} \right)\,,
\label{K}
 \ee
and a polynomial formulae
 \be {\mathcal{FF}}^*=\sum_{k=0}^n a_k H^k,
 \qquad {\mathcal{F}}^*{\mathcal{F}}=\sum_{k=0}^n a_k^* (H^*)^k\,
\label{Poly}
 \ee
with the final goal to elucidate the conditions leading to such a
type of the closure of the algebra~\cite{Sokolovdva}.

\subsection{Plan of the paper}

In Section \ref{SSUSY} we elaborate a particular solution to our
problem inspired by the specific second-order supersymmetry
(SSUSY) results of ref.~\cite{Ba02}. Our solution of Eq.
(\ref{Poly}) will have the form
 \be {\mathcal{FF}}^*=h_1^2-\frac{c^2}{4},
  \qquad {\mathcal{F}}^*{\mathcal{F}}=h_2^2-\frac{c^2}{4}\,,
 \ee
where $h_1$ is naturally related to $h_2$ by Hermitian
conjugation,
 \be h_1=h_2^\dag\,,
\label{Herm_C}
 \ee
if $c^2$ is real. Our explicit solution to the problem is rendered
possible by a SSUSY inspired gluing constraint
\cite{Sokolov,Sokolovdva}. We show that
 \be {\mathcal{F}} h_2 = h_1{\mathcal{F}}\,,
 \ee
which, because of Eq. (\ref{Herm_C}), is now equivalent to Eq.
(\ref{inter_rel}). This amounts to
 \be {\mathcal{CPT}} h_1= h_1{\mathcal{CPT}}\,.
\label{CPT_cons}
 \ee
Explicit analytic examples of ${\cal PT}-$asymmetric models are
expressed in terms of circular or hyperbolic functions.

In Section \ref{SCO} we perform a detailed investigation of eq.
(\ref{inter_rel}) for a charge operator which is of the second
order in derivatives,
 \begin{equation}
  {\mathcal{C}} = \frac{d^2}{dx^2}+G\left( x\right)
  \frac d{dx} +D\left( x\right) ,
\label{charge}
 \end{equation}
and where $G\left( x\right) $ and $D\left( x\right) $ are complex
functions of the real coordinate $x$:
 \begin{eqnarray}
 G\left( x\right) &=&G_R\left( x\right) +iG_I\left( x\right) ,
 \nonumber \\ D\left( x\right) &
 =&D_R\left( x\right) +iD_I\left( x\right).
 \nonumber
 \end{eqnarray}
We further derive the polynomial algebra of Eq. (\ref{Poly}). In
order to show explicitly that our formalism allows to generate
${\cal PT}-$asymmetric models with real spectrum, we discuss in
Section \ref{part} a particular polynomial oscillator model.

In Section \ref{Charge} we generalize the postulate (\ref{charge})
and derive the general form of the charge operator $\mathcal{C}$
of any finite order in the derivative $d/dx$ such that
$\mathcal{F}\equiv\mathcal{CP}$ is Hermitian. At the very end, in
section \ref{five} we give some perspectives on the impact of our
results on a variety of fields where the use of similar
$\mathcal{F}$ might play significant role.

\section{SUSY gluing constraint
\label{bi}}

% \section{An Algebraic Approach}
\label{SSUSY}

Starting with a second-order $\mathcal{C}$ of the form
(\ref{charge}) we have to guarantee, first of all, the Hermiticity
of ${\mathcal{F}}={\cal C}{\cal P}$ and ${\mathcal{F}}^{-1}={\cal
P}{\cal C}^{-1}$. It is easy to show (see also section \ref{ego}
below for an exhaustive discussion of these important conditions
for polynomial charges) that the latter Hermiticity condition
forces us to impose the necessary and sufficient requirements
 \ben
 D_R(x) = D_R(-x) + \frac{d}{dx}G_R(x), \ \ \ \ \ \ \
 D_I(x) = -D_I(-x) + \frac{d}{dx}G_I(x)\,
 \een
where $G_R\left( x\right)=G_R\left(- x\right)$ is even while
$G_I\left( x\right)=-G_I\left(- x\right)$ must be odd.

\subsection{Factorization}

In the subsequent step of our considerations we factorize our
second-order charge operator $\mathcal{C}$ as follows,
  \be
   {\mathcal{C}} = q_1q_2\,,\quad q_1 = \frac{d}{dx}+U(x)\,,
 \quad q_2 = \frac{d}{dx}+W(x)\,,
\label{q_1q_2}
  \ee
where
  \be U(x)+W(x)=G(x)\,,\qquad
\frac{d}{dx}W(x)+U(x)W(x)=D(x)\,. \label{UW}
  \ee
In order to simplify the problem at the start we impose the
following ``gluing" constraint on $q_1$ and $q_2$,
 \be
  q_2(q_2^{\dag})^*=(q_1^{\dag})^*q_1+c\,, \label{constraint}
 \ee
where $c$ is a complex number. By inserting Eqs. (\ref{q_1q_2})
into Eq. (\ref{constraint}), we obtain
 $$
 \left( \frac{d}{dx}+W\right)\left(-\frac{d}{dx}+W\right)
 =\left(-\frac{d}{dx} +U\right)\left(\frac{d}{dx}+U\right)+c\,,
 $$ whence
 \be
\frac{d}{dx} W(x)+W^2(x)=-\frac{d}{dx}U(x)+U^2(x)+c\,.
\label{UW_1}
 \ee
We find the following representation for ${\mathcal{FF}}^*$ and
${\mathcal{F}}^*{\mathcal{F}}$ (Eq. (\ref{F}))
 $$
 {\mathcal{FF}}^*={\mathcal{F}}\left({\mathcal{F}}^\dag\right)^*
 =\left(q_1q_2{\mathcal{P}}\right)\cdot
                 \left({\mathcal{P}}q_2^\dag q_1^\dag\right)^*
                 =q_1q_2\left({\mathcal{P}}\right)^2
                 \left(q_2^\dag\right)^*\left(q_1^\dag\right)^*\,,
 $$
which, taking Eq. (\ref{constraint}) into account, becomes
 \begin{eqnarray*} {\mathcal{FF}}^*
  &=& q_1\left[\left(q_1^\dag\right)^*q_1
+\frac{c}{2}+\frac{c}{2}\right]\left(q_1^\dag\right)^* \\
                 &=& \left[ q_1\left(q_1^\dag\right)^*
                 +\frac{c}{2}+\frac{c}{2}\right]\cdot
                     \left[q_1\left(q_1^\dag\right)^*
                     +\frac{c}{2}-\frac{c}{2}\right]\,.
 \end{eqnarray*}
Correspondingly,
 \begin{eqnarray*} {\mathcal{F}}^*{\mathcal{F}}
  &=& \left({\mathcal{F}}^\dag
\right)^*{\mathcal{F}}\\
                             &=& \left[{\mathcal{P}}
                      \left(q_2^\dag \right)^*q_2{\mathcal{P}}
                             -\frac{c}{2}-\frac{c}{2}\right]
                                 \cdot \left[{\mathcal{P}}
                                 \left(q_2^\dag \right)^*
                                  q_2{\mathcal{P}}-\frac{c}{2}
                                  +\frac{c}{2}\right]\,.
 \end{eqnarray*}
Defining the Hamiltonian operators
 \begin{eqnarray} h_1 &=& q_1(q_1^{\dag})^*
 +\frac{c}{2}\nonumber \\
    &=& \left( \frac{d}{dx}+U\right)
    \left(-\frac{d}{dx}+U\right)+\frac{c}{2}\nonumber \\
    &=& -\frac{d^2}{dx^2}+\frac{d}{dx}U(x)+U^2(x)
    +\frac{c}{2}\nonumber \,,
 \end{eqnarray}
and
 \begin{eqnarray}
 h_2 &=& {\mathcal{P}}(q_2^{\dag})^* q_2{\mathcal{P}}
 -\frac{c}{2}\nonumber \\
    &=& {\mathcal{P}}\left(-\frac{d}{dx}+W(x)\right)
    \left(\frac{d}{dx}+W(x)\right){\mathcal{P}}-\frac{c}{2}\nonumber \\
    &=& -\frac{d^2}{dx^2}-\frac{d}{dx}W(-x)+W^2(-x)
    -\frac{c}{2}\nonumber\,,
 \end{eqnarray}
equation (\ref{K}) provides the following representation for $K$
 $$ K={\cal H}^2-\frac{c^2}{4}\,,
\ \ \ \ \ \  {\cal H}=\left(
 \begin{array}{cc} h_1 & 0 \\ 0   & h_2
 \end{array}
\right)\,.
 $$
Comparing with Section \ref{poly} below, where ${\cal F
F}^*=H^2+\alpha H +\gamma$, and setting $h_1 = H$, we get in the
present case $\alpha=0$ and, correspondingly, $V_0=0$, according
to Eq. (\ref{Alpha}) below, as well as $\gamma=-c^2/4$. This shows
explicitly how the present model can be derived from the general
results of section \ref{SCOPE}.

\subsection{Hamiltonians}

Remembering the first of Eqs. (\ref{UW}), Eq. (\ref{UW_1}) becomes
 $$
\frac{d}{dx}G(x)+W^2(x)-(G(x)-W(x))^2=c\,,
 $$
\textit{i.e.},
 $$
\frac{d}{dx}G(x)-G^2(x)+2G(x)W(x)=c\,,
 $$ or
 \be G(x)W(x)=\frac{1}{2}\left(G^2(x)-\frac{d}{dx}G(x)+c\right)\,.
\label{GW}
 \ee
We immediately deduce that
 \begin{eqnarray} W(x) &=& \frac{G^2(x)-\frac{d}{dx}G(x)+c}{2G(x)}
  \,,\nonumber \\ U(x) &=&
G(x)-W(x)=\frac{G^2(x)+\frac{d}{dx}G(x)-c}{2G(x)} \,. \label{U}
 \end{eqnarray} Thus
 $$ h_1=-\frac{d^2}{dx^2}+V(x)\,,
 $$ with
 \be V(x)=G'(x)-\frac{(G'(x))^2}{4G^2(x)}+\frac{G''(x)}{2G(x)}
 +\frac{G^2(x)}{4}
     +\frac{c^2}{4G^2(x)}\,.
\label{V_x}
 \ee
From Eq. (\ref{GW}) we also get that at the zeros $\bar x$ of $G$,
we must have
 $$
\frac{dG}{dx}\bigg|_{x=\bar x}=c\,, $$ which is a constraint on
$G$, too. In fact, the method would fail if $G$ had several zeros
with non-identical values of the first derivative at each of them.

An important comment must be made here since even if a function
does not vanish on the real axis, one can investigate its zeros in
the complex $x$ plane. For instance, if
 \be G(x)=G_0(x)\equiv z(x)=\frac{1+i\sinh(\alpha x)}{2}\,,
 \quad \alpha\in\mathbf{R},
 \label{cioffi}
 \ee
it is immediate to check that $z(x_n)=0$ at
$x_n=-i\,(2n+3/2)\pi/\alpha$, $n=0$, $\pm 1$,~\ldots. This would
mean that $dG_0(x_n)/dx=(i\alpha/2)\cosh(\alpha x_n)=0$, thus
implying that we must put $c=0$ in this case.

In the similar spirit, we may consider the whole class of
functions which depend on $x$ only via $z(x)$ of Eq.
(\ref{cioffi}) in an arbitrary nonlinear manner, $G_m(x)\equiv
G(z(x))$, since, as a function of $x$, $z$ is
$\mathcal{PT}$-symmetric, and any real function of $z$ is
$\mathcal{PT}$-symmetric, too, and is an acceptable candidate for
$G$.

It becomes convenient to change variables and express the
Hamiltonian, $H=-d^2/dx^2+V(x)$, with $V(x)$ given by formula
(\ref{V_x}), as a function of $z$, by observing that
 \begin{eqnarray*}
\frac{d}{dx} &=&\frac{dz}{dx}\frac{d}{dz}=i\alpha\sqrt{z(1-z)}
\frac{d}{dz}\,, \\ \frac{d^2}{dx^2}
&=&\left(\frac{dz}{dx}\frac{d}{dz}\right)^2
 =-\alpha^2\left(\frac{1}{2}-z\right)\frac{d}{dz}
                   -\alpha^2z(1-z)\frac{d^2}{dz^2}\,,
 \end{eqnarray*} and
 \begin{eqnarray} V(z)&=&i\alpha\sqrt{z(1-z)}\frac{d}{dz}G+\alpha^2
 \frac{z(1-z)}{4G^2}\left(\frac{d}{dz}G\right)^2
     -\alpha^2\frac{1-2z}{4G}\frac{d}{dz}G \label{V_z}\\
    & &-\alpha^2\frac{z(1-z)}{2G}\frac{d^2}{dz^2}G +\frac{G^2}{4}
    +\frac{c^2}{4G^2}\,.\nonumber
 \end{eqnarray}

\subsection{Consistency}

We prove now an important constraint on the complex number
$c=c_R+ic_I$. From the second of Eqs. (\ref{UW}), we have
 \begin{eqnarray} D(x) &=& \frac{d}{dx}W(x)+U(x)W(x) \nonumber \\
     &=&\frac{1}{2G^2(x)}\left[\left(2G\frac{d}{dx}G
     -\frac{d^2}{dx^2}G\right)G
       -\frac{d}{dx}G\left(G^2-\frac{d}{dx}G+c\right)\right]
        \nonumber \\
     & &+\frac{1}{4G^2}\left[G^4-\left(\frac{d}{dx}G\right)^2
     -c^2+2c \frac{d}{dx}G\right]\,, \nonumber
 \end{eqnarray} or
 \begin{eqnarray} D(x) &=& \frac{1}{4G^2}\left[2G^2\frac{d}{dx}G
-2G\frac{d^2}{dx^2}G+\left(\frac{d}{dx}G\right)^2
       +G^4-c^2\right] \,, \label{D_Em}\\ D^*(-x)
        &=& \frac{1}{4G^2}\left[-2G^2\frac{d}{dx}G
-2G\frac{d^2}{dx^2}G+\left(\frac{d}{dx}G\right)^2
       +G^4-(c^*)^2\right] \,, \label{D*_Em}
 \end{eqnarray}
where the functions on the right-hand-sides of Eqs. (\ref{D_Em})
and (\ref{D*_Em}) are all computed at $x$.

In deriving Eq. (\ref{D*_Em}), use has been made of the fact that
$G$ and $d^2G/dx^2$ are $\mathcal{PT}$-symmetric, while $dG/dx$ is
$\mathcal{PT}$-antisymmetric, \textit{i.e.},
$(dG/dx(-x))^*=-dG/dx(x)$. Eq. (\ref{D_Em}) is obviously
consistent with the general form of $D$ as a function of $G$ given
by Eqs. (\ref{D_f}), (\ref{I_0}), with
$c^2/4=-I_0-D(x_0)G^2(x_0)$. Subtracting Eq. (\ref{D*_Em}) from
Eq. (\ref{D_Em}) side by side, we obtain
 \be D(x)-D^*(-x)=\frac{d}{dx}G+\frac{(c^2)^*-c^2}{4G^2}\,.
\label{D-D*_1}
 \ee
Combining Eq. (\ref{D-D*_1}) with Eq. (\ref{D-D*}), we obtain the
important result
 $$(c^2)^*-c^2=0\quad \rightarrow \quad \Im (c^2)=0
 \quad \rightarrow \quad c_Rc_I=0.
   $$ From Eq. (\ref{U}) we easily obtain
 $$ U(x)=W^*(-x)-\frac{c_R}{G(x)}\,,  $$ whence
 \be
\left(\frac{d}{dx}U(x)\right)^*=-\frac{d}{dx}W(-x)
 +\frac{c_R}{(G^*(x))^2}\frac{d}{dx}G^*\,,
 \ee
and
 \be
\left(U^*(x)\right)^2=W^2(-x)+\frac{c_R^2}{(G^*(x))^2}
 -2c_R\frac{W(-x)}{G^*(x)}\,.
 \ee
Thus
 \begin{eqnarray} h_1^\dag -h_2 &=& \left(\frac{d}{dx}U(x)
+U^2(x)\right)^*+\frac{d}{dx}W(-x)-W^2(-x)+c_R \nonumber \\
              &=& c_R\left[\frac{1}{(G^2(x))^*}
              \left(\frac{d}{dx}G^*(x)+c_R\right)
              -2\frac{W(-x)}{G^*(x)}+1\right] \,.\nonumber
 \end{eqnarray} Using Eq. (\ref{U}) to replace $W(-x)$, we obtain
 \begin{eqnarray} h_1^\dag -h_2 &=&c_R\left[\frac{1}{(G^2)^*}
\left(\frac{d}{dx}G^*+c_R\right)-\frac{1}{(G^2)^*}\left((G^2)^*
 +\frac{d}{dx}G^*+c\right)
                 +1\right] \nonumber \\
              &=& c_R \frac{c_R-c}{(G^2)^*}
              =-i\frac{c_R c_I}{(G^2)^*}=-ic_R c_I
               \frac{G^2}{\mid G\mid ^4}\,.
 \end{eqnarray} Therefore
 $$ h_1^\dag = h_2 \quad \Leftrightarrow \quad c_R c_I = 0 \,.  $$

\subsection{Periodic potential }

Let us now give an example which generalizes
$\mathcal{PT}$-symmetric periodic potentials~\cite{Ca98,Ah01}:
 $$ G(x) = e^{i\alpha x}+r\,, \quad\alpha\in{\mathbf R}\,,
 \ \ \quad r \in{\mathbf R}, \ \ r \neq \pm 1.
 $$
 In this case we have, for all $ x\in\mathbf R$,
 \ben
 U(x) = \frac{1}{2}\left (
 e^{i\alpha x}+r\right)  +
 \frac{1}{2}\left (i\alpha e^{i\alpha
 x}-c\,\right )/
 \left (
 e^{i\alpha x}+r\right)
 \een
 \ben
 W(x) = \frac{1}{2}\left (
 e^{i\alpha x}+r\right)  -
 \frac{1}{2}\left (i\alpha e^{i\alpha
 x}-c\,\right )/
 \left (
 e^{i\alpha x}+r\right)\,.
 \een
Since $G$ never vanishes, we do not have any constraint on the
value of $c$, in addition to the one which requires that $c$ be
either real, or imaginary. The spectral analysis of the
corresponding Schr\"odinger operators $h_1$ and $h_2$ with
periodic potentials can be performed as a generalization to the
non-$\mathcal {PT}$-symmetric case of the investigation done by in
Ref.~\cite{Sh02}.

We now examine the invertibility of ${\cal C }$ and the
boundedness of  ${\cal C }^{-1}$. First notice that ${\cal C }$
can be written in the following form
 \be
 {\mathcal{C}}={\mathcal{C}}_1{\mathcal{C}}_2,
 \
 \
 \ \ \ \ \
 \
 {\mathcal{C}}_1=
  {\mathcal{C}}_U+\frac{r}{2}, \ \ \ \ \ \ \ \
  {\mathcal{C}}_2=
  {\mathcal{C}}_U+\frac{r}{2}
 \label{faktori}
 \ee
where
 \ben
 {\mathcal{C}}_U=\frac{d}{dx}+U_1, \ \ \ \ \
 U_1 = U-\frac{r}{2}
 \een
 \ben
 {\mathcal{C}}_W=\frac{d}{dx}+W_1, \ \ \ \ \
 W_1 = W-\frac{r}{2}\,.
 \een
We will discuss the invertibility of each factor in
(\ref{faktori}) separately. As for ${\mathcal{C}}_1$ we first
observe that the numerical range $\{z=\langle {\mathcal{C}}_U
\psi,\psi\rangle : \psi \in H^{1}(I\!\!R)\}$ of ${\mathcal{C}}_U$
is contained in the strip $\{z : |Re\,z| \leq a\}$ where $a =
{\max}_{x \in I\!\!R }|U_1(x)|$. Hence, if $|r| > 2a$ then $-r/2$
is in the resolvent set of ${\mathcal{C}}_U$ \cite{Kato} and,
therefore, ${\mathcal{C}}_1$ is invertible with bounded inverse on
$L^2 (\mathbf R)$. A similar argument holds for ${\mathcal{C}}_W$.
Thus, for sufficiently large values of $|r|$, operator ${\cal C }$
is invertible and ${\cal C }^{{-1}}$ is bounded on $L^2 (\mathbf
R)$.

\section{Second-order charge operator ${\cal
C}$
\label{SCOPE}}
 \label{SCO}

\subsection{Re-construction of the potential}

 %Intertwining relations}
 % with a Metric Operator \\ Involving SCO}

We already noticed that in the second-order charge operator
(\ref{charge}), the notation of Section \ref{Charge} below implies
that we have the correspondences $\gamma_2(x)=1$, $\gamma_1
(x)=G(x)$ and $\gamma_0 (x)=D(x)$, so that the Hermiticity
constraints on the real and imaginary parts of $\gamma_{\ell}(x)$,
Eqs. (\ref{Reca}) and (\ref{recb}), with $\omega =2$ and $\ell
=0,1$, immediately give
 \begin{eqnarray} G_R\left( x\right)-G_R\left( -x\right) &=& 0\, ;
 \quad G_I\left(x\right) +G_I\left( -x\right)=0 ;
  \label{G} \\ D_R\left( x\right)-D_R\left( -x\right) &=& \frac
  d{dx}G_R\left(
  x\right)
   ;\quad D_I\left(
x\right) +D_I\left( -x\right) = \frac d{dx}G_I\left(
x\right)\,.\label{D_0}
 \end{eqnarray}
As a consequence of Eq. (\ref{G}), $G$ is
 $\mathcal{PT}$-symmetric
 \be G(x)=G^*(-x)\,,
\label{G-G*}
 \ee
while Eq. (\ref{D_0}) yields
 \be D(x)-D^*(-x)=\frac{d}{dx}G(x)\,.
\label{D-D*}
 \ee
We assume that $\mathcal{F}$ and  $H$  satisfy the intertwining
condition (\ref{inter_rel}) and that $H$ depends on a local
complex potential, $V\left( x\right) $:
 \begin{equation}
  H=-\frac{d^2}{dx^2}+V\left( x\right) ,  \label{hamiltonian}
 \end{equation}
with $V\left( x\right) =V_R\left( x\right) +iV_I\left( x\right)
  $. In turn, $V_R\left( x\right)
$ and $V_I\left( x\right) $ are conveniently decomposed into their
even and odd parts: $_{}$
 \begin{eqnarray*} V_R\left( x\right) &=&V_R^E\left( x\right)
  +V_R^O\left(
 x\right )
  , \\ V_I\left( x\right)
&=&V_I^E\left( x\right) +V_I^O\left( x\right) ,
 \end{eqnarray*}
with $V_K^E\left( x\right) =V_K^E\left( -x\right) $ and
$V_K^O\left( x\right) =-V_K^O\left( -x\right) $, ($K=R,I)$. We
write now condition $\left( \ref{inter_rel}\right) $ explicitly
and obtain three non-trivial equations by imposing that the
coefficients of $(d/dx)^2$, $d/dx$ and $(d/dx)^0$ vanish,
 \be
 -2\left( V_R^O +iV_I^E \right)
  +2\frac d{dx}\left( G_R+iG_I \right) = 0\,,
 \label{a=0}
 \ee
 \be
 \begin{array}{l}
 2\frac
 d{dx}\left(
 V_R^E +iV_I^O \right)
 -2\frac d{dx}\left( V_R^O +iV_I^E \right)
+2\frac d{dx}\left( D_R +iD_I
 \right)
\label{b=0}\\
 +\frac{d^2}{dx^2}\left( G_R +iG_I \right)
  -2\left( V_R^O +iV_I^E \right)
  \left( G_R +iG_I \right)=0 \,,
  \ea
  \ee
  \be
 \begin{array}{l}
  \frac{d^2}{dx^2}\left( V_R^E +iV_I^O \right)
-\frac{d^2}{dx^2}\left( V_R^O +iV_I^E \right) +
\frac{d^2}{dx^2}\left( D_R +iD_I \right)\\ +\left( G_R +iG_I
\right) \frac d{dx}\left( V_R^E\left( x\right) +iV_I^O\left(
x\right) \right) -\left( G_R
 +iG_I \right) \frac d{dx}\left( V_R^O +iV_I^E \right)
  \\ -2\left( D_R +iD_I \right) \left( V_R^O +iV_I^E
\right)=0 \label{c=0}
 \end{array}
 \ee
while
the coefficients of $(d/dx)^4$ and $(d/dx)^3$ are identically
zero.

\subsection{Integrability}

The first of the above equations (\ref{a=0}) yields
 \begin{equation} V_R^O  =\frac d{dx}G_R ;
  \qquad V_I^E =\frac d{dx}G_I\, .\nonumber \label{FP}
\label{V_1}
 \end{equation}
The second equation  (\ref{b=0})yields
 \begin{eqnarray}
\frac d{dx}V_R^E -\frac d{dx}V_R^O -V_R^O G_R +V_I^E G_I +\frac
d{dx}D_R +\frac 12\frac{d^2}{dx^2}G_R  &=&0;\nonumber \\ \frac
d{dx}V_I^O -\frac d{dx}V_I^E -V_I^E G_R -V_R^O
 G_I +\frac d{dx}D_I +\frac 12\frac{d^2}{dx^2}G_I
&=&0,\nonumber
 \end{eqnarray}
and is easily integrated for the other two components of the
potential, $V_R^E\left( x\right) $ and $V_I^O\left( x\right) $, as
functions of $ G_R\left( x\right) $, $G_I\left( x\right) $,
$D_R\left( x\right) $ and $ D_I\left( x\right) $, by replacing
$V_R^O\left( x\right) $ and $V_I^E\left( x\right) $ with their
expressions $\left( \ref{FP}\right) $:
 \begin{eqnarray} V_R^E\left( x\right)
\!\!&\!=\!&\!\!\frac 12\frac d{dx}G_R\left( x\right) +\frac
12\left( G_R\left( x\right) \right) ^2\!-\frac 12\left( G_I\left(
x\right) \right) ^2\!-D_R\left( x\right) +V_0
  \label{SP}  \\ V_I^O\left( x\right)\!\!&\!\!=&\!\!\frac 12\frac
d{dx}G_I\left( x\right) + G_R\left( x\right) G_I\left(
 x\right) -D_I\left( x\right) .
\label{V_2}
 \end{eqnarray}
Here, $V_0 $  is a real  integration constant.
The corresponding integration constant in the equation for
$V_I^O(x)$ must be zero, because the function is odd. Both
equations can be recombined as
 \be
  V(x)=\frac{3}{2}\frac{d}{dx}G(x)+\frac{1}{2}G^2(x)-D(x)+V_0\,.
\label{V}
 \ee
Finally, the third equation  (\ref{c=0}) allows us
to express the $G_J\left( x\right) $'s, $\left( J=R,I\right) $, as
functions of the $D_K\left( x\right) $'s, $\left( K=R,I\right) $,
or, more conveniently, viceversa.
 \begin{eqnarray} &&-\frac 12\frac{d^3}{dx^3}G_R
 +\frac{G_R}2\frac{d^2}{dx^2}G_R+\left(
 \frac
d{dx}G_R\right) ^2+\left( G_R^2-G_I^2 -2D_R\right) \frac d{dx}G_R
\label{TP} \nonumber \\ &&-\frac{G_I}2\frac{d^2G_I}{dx^2} -\left(
\frac d{dx}G_I\right) ^2+2\left( D_I-G_IG_R\right) \frac d{dx}G_I
-G_R\frac d{dx}D_R+G_I\frac d{dx}D_I\nonumber \\ &=&0 \\ &&-\frac
12\frac{d^3G_I}{dx^3} +\frac 12G_R\frac{d^2}{dx^2}G_I+\left(
-G_I^2+G_R^2-2D_R\right) \frac d{dx}G_I +\frac
12G_I\frac{d^2}{dx^2}G_R \nonumber \\ &&+ 2\frac d{dx}G_R  \frac
d{dx}G_I +2(G_RG_I -D_I)\frac d{dx}G_R-G_I\frac d{dx}D_R-G_R\frac
d{dx}D_I \nonumber \\ &=&0.\nonumber
 \end{eqnarray}
Eqs. $\left( \ref{TP}\right) $ can be recombined in the following
first-order linear equation expressing the unknown function $D(x)$
in terms of the known function $G(x)$ and its derivatives up to
third order
 \be
\frac{1}{2}\frac{d^3}{dx^3}G-\frac{1}{2}G\frac{d^2}{dx^2}G
 -\left(\frac{d}{dx}G\right)^2
 -G^2\frac{d}{dx}G+2\left(\frac{d}{dx}G\right)D+G\frac{d}{dx}D=0\,.
\label{D_G}
 \ee
Eq. (\ref{D_G}) is easily solved by direct integration.  Let us
define the auxiliary functions
 \begin{eqnarray} g(x) &\equiv& 2\frac{d}{dx}G\,, \label{g}
  \\ f(x) &\equiv&
-\frac{1}{2}\frac{d^3}{dx^3}G+\frac{1}{2}G\frac{d^2}{dx^2}G
+\left(\frac{d}{dx}G\right)^2
         +G^2\frac{d}{dx}G\,,\\
\frac{1}{p(x)}\frac{d}{dx}p \label{p}(x)&\equiv &
 \frac{g(x)}{G(x)}\,.
 \end{eqnarray}
Eq. (\ref{p}) is promptly integrated by use of definition
(\ref{g}) to
 \be
 p(x)\,=\,\exp\left(2\int_{x_0}^x d \ln
 G(x')\right)\,=\,\frac{G^2(x)}{G^2(x_0)}\,,
 \ee
where $x_0$ is an initial point where $G$ is different from zero.
It is now easy to check that the general solution to Eq.
(\ref{D_G}) can be written in the form
 $$ p(x)D(x)
  = \int_{x_0}^x dx'\frac{p(x')f(x')}{G(x')}+p(x_0)D(x_0)\,,
 $$ or
 \be
 D(x)= \frac{1}{G^2(x)}\int_{x_0}^x dx' G(x')f(x')
 +\frac{D(x_0)G^2(x_0)}{G^2(x)}\,. \nonumber
\label{D_f}
 \ee
The integral on the right-hand side of Eq. (\ref{D_f}) is computed
by elementary  methods in the form
 \begin{equation}
\int_{x_0}^x dx' G(x')f(x') =
\frac{G^4(x)}{4}+\frac{G^2(x)G'(x)}{2}
-\frac{G(x)G''(x)}{2}+\frac{(G'(x))^2}{4}+I_0 \,, \label{int}
 \end{equation} with
 \be I_0\equiv - \frac{G^4(x_0)}{4}-\frac{G^2(x_0)G'(x_0)}{2}
+\frac{G(x_0)G''(x_0)}{2}-\frac{(G'(x_0))^2}{4}  \,, \label{I_0}
 \ee
where $G'\equiv dG/dx$, and so on, thus providing the most general
expression of $D$ as a function of $G$ and of its derivatives.

\subsection{SSUSY algebra}
\label{poly}

Assuming a charge operator, ${\mathcal{C}}(x)$, of the form
(\ref{charge}), we now verify that the operator
 $${\mathcal{F}}(x){\mathcal{F}}^*(x)
 ={\mathcal{C}}(x){\mathcal{PC}}^*(x){\mathcal{P}}={\mathcal{C}}(x)
  {\mathcal{C}}^*(-x)
 $$ can be written as a particular case of formula (\ref{Poly})
 $$ {\mathcal{F}}(x){\mathcal{F}}^*(x)=H^2+\alpha H+\gamma\,,
 $$ where $\alpha$ and $\gamma$ are constants to be determined and
$H$ is Hamiltonian (\ref{hamiltonian}) with $V$ given in
(\ref{V}). In fact, we have
 \begin{eqnarray*}
 {\mathcal{C}}(x){\mathcal{C}}^*(-x)\!&\!=\!&\left(\frac{d^2}{dx^2}
+G(x)\frac{d}{dx}+D(x)\right)\cdot
                                      \left(\frac{d^2}{dx^2}
                               -G^*(-x)\frac{d}{dx}+D^*(-x)\right) \\
                                   &\!=\!&\left(\frac{d^2}{dx^2}
                                   +G(x)\frac{d}{dx}+D(x)\right)\cdot
                                      \left(\frac{d^2}{dx^2}
                                  \!-\!G(x)\frac{d}{dx}+D(x)
                                  \!-\!G'(x)\right)
 \end{eqnarray*}
where use has been made of relations (\ref{G}), (\ref{D_0})
stemming from Hermiticity of $C(x)$. After some algebra, the
right-hand side of the above expression is brought to the form
 \begin{eqnarray} {\mathcal{C}}(x){\mathcal{C}}^*(-x)
                                   &=&\frac{d^4}{dx^4}
                                   +\left(2D(x)-G^2(x)
                                   -3G'(x)\right)\frac{d^2}{dx^2}\\
                                   & &+\left(2D'(x)-3G''(x)
                                   -2G(x)G'(x)\right)\frac{d}{dx}
                                    \nonumber\\
                                   & &+D''(x)-G'''(x)+G(x)D'(x)
                                    \nonumber \\
                                   & &-G(x)G''(x)+D^2(x)
                                   -D(x)G'(x)\,,\nonumber
 \end{eqnarray} and is to be compared with
 \begin{eqnarray} H^2+\alpha H+\gamma
                    &=& \left(-\frac{d^2}{dx^2}+V(x)\right)^2
                    +\alpha\left(-\frac{d^2}{dx^2}+V(x)\right)
                    +\gamma\\
                    &=&\frac{d^4}{dx^4}-\left(2V(x)+\alpha\right)
                    \frac{d^2}{dx^2}-2V'(x)\frac{d}{dx}+V^2(x)
                     \nonumber \\
                    & & -V''(x)+\alpha V(x)+\gamma \,,\nonumber
 \end{eqnarray}
where $V(x)$ may be expressed as a function of $D(x)$ and $G(x)$
according to Eq. (\ref{V}). Direct comparison of the right-hand
sides of the above formulae allows us to determine the $\alpha$
constant as
 \be
\alpha=-2V_0\,. \label{Alpha}
 \ee
The value of $\gamma$ expresses the compatibility between ${\cal
C}$ and the polynomial algebra through the equation
 $$
 V^2(x) -V''(x)+\alpha V(x)+\gamma=
  $$
  $$
  =D''(x)-G'''(x)+G(x)D'(x)
 -G(x)G''(x)+D^2(x)-D(x)G'(x).
 $$ Here, we insert the expressions of $V(x)$ and $V''(x)$ in terms of
$G(x)$, $D(x)$ and of their derivatives obtained from formula
(\ref{V}), and making use of Eq. (\ref{D_G}), as well as of its
general solution (\ref{D_f}), (\ref{int}), we obtain the final
result
 \be
  \gamma = V_0^2+I_0+D(x_0)G^2(x_0)\,, \label{Gamma}
  \ee
where $I_0$ is defined in Eq. (\ref{I_0}). This makes it possible
to interpret $\gamma$ as a kind of integration constant. Thus,
$\mathcal{CPT}$ invariance leads to the SSUSY polynomial algebra,
Eqs. (\ref{K}), (\ref{Poly}).

\section{Polynomial oscillators
\label{Ema}}

%\section{The Simplest SCO Model}
\label{part}

The simplest factorization of  ${\cal C}$ reads
 \be {\mathcal{C}}(x)=\left(\frac{d}{dx}+\frac{G(x)}{2}\right)
\cdot\left(\frac{d}{dx}+\frac{G(x)}{2}\right)\,,
\label{factorization}
 \ee
so that, correspondingly,
 \be D(x)=\frac{G'(x)}{2}+\frac{G^2(x)}{4}\,.
 \ee
In this case, Eq. (\ref{D_G}) yields $G'''(x)=0$, \textit{i.e.},
 \be \label{eq6} G(x)=ax^2 +ibx+c
 \ee
where $a,b$ and $c$ are real numbers, owing to the fact that
$G(x)$ is $\mathcal{PT}$-symmetric. From Eq. (\ref{V})  we obtain:
  \begin{eqnarray}
\lefteqn{ V(x) = \frac{1}{4}G^2 (x)+G'(x)+V_0 {} }
 \nonumber \\ \!\!&\!\! & \!\! =\frac{1}{4}a^2 x^4\!
-\frac{1}{4}(b^2\!-2ac)x^2\!
 +\frac{1}{2}iabx^3\! +\frac{1}{2}x(ibc+4a)+ib+\frac{c^2}{4}+V_0.
  \end{eqnarray}
If we make the additional assumption $c=0$, for the sake of
simplicity, the polynomial algebra provides the constraint
 $$
\gamma=V_0^2+\frac{b^2}{4}\,
 $$
on $\gamma$ [Eq. (\ref{Gamma})].

\subsection{The problem of invertibility}

We will now make the spectral analysis for $H$ and study the
invertibility of ${\mathcal F}$ in the case $c=0$. Then
 \be V(x)=\frac{1}{4}a^2 x^4 -\frac{1}{4}b^2 x^2 +\frac{1}{2}iabx^3
  +2ax+ib+V_0.
 \ee
Setting $\mu ^2 =\frac{a^2}{4}$ and $\nu ^2 =\frac{b^2}{4}$, we
obtain an expression for the Schr\"odinger operator $H$ of the
same type as that presented in Eqs. (22), (23) of
Ref.~\cite{Ca04}, namely
 \be
  H=-\frac{d^2}{dx^2}+\mu ^2 x^4 -\nu ^2 x^2 +2i\mu
\nu x^3 +4\mu x+2i\nu +V_0
 \label{63}
 \ee
and $D(H)=H^2 ({\bf R})\cap D(x^4),\;\forall \mu,\nu \in {\bf R
},\,\mu \ne 0$. As in Ref.~\cite{Ca04}, $H$ has discrete spectrum,
\textit{i.e.}, the spectrum consists of a sequence of isolated
eigenvalues with finite multiplicity.

In order to prove the reality of the spectrum of $H$, we first
notice that $H$ can be rewritten as
 \be H=-\frac{d^2}{dx^2}+x^2 (\mu x+i\nu)^2 +4\mu x+2i\nu +V_0.
 \ee
Let us now perform the complex translation $x\to
x-\frac{i\nu}{2\mu}$. Then $H=S^{-1}H_1 S$ where $S\psi (x)=\psi
(x-\frac{i\nu}{2\mu})$ on a dense set of functions $\psi \in L^2
(\mathbf R)$ and
 \begin{eqnarray} H_1 & =& -\frac{d^2}{dx^2}
 +\left(x-\frac{i\nu}{2\mu}\right)^2
 \left(\mu x-\frac{i\nu}{2}+i\nu \right)^2
  +4\mu x-2i\nu +2i\nu +V_0 {}
\nonumber
\\ & & {} =-\frac{d^2}{dx^2}+\mu ^2
\left(x-\frac{i\nu}{2\mu}\right)^2
\left(x+\frac{i\nu}{2\mu}\right)^2 +4\mu x+V_0 \nonumber
 \\ & & {} =-\frac{d^2}{dx^2}+\mu ^2
\left(x^2 +\frac{\nu ^2}{4\mu ^2}\right)^2 +4\mu x+V_0
 \end{eqnarray}
Hence $H$ has the same spectrum of $H_1$. In turn $H_1$ is
selfadjoint on $D(H_1)=D(H)=H^2 ({\bf R})\cap D(x^4)$, thus it has
real spectrum for all $\mu,\nu,V_0 \in {\bf R},\, \mu \ne 0$.

We may stress that Hamiltonian (\ref{63}) is not ${\cal PT}-$invariant but
has still a real spectrum because it is related by explicit similarity
to the standard self-adjoint anharmonic oscillator. In our opinion
this is an exceptional example since in general the proof of the
reality of the spectra of non-Hermitian Hamiltonians cannot proceed in
such a straightforward manner and, generically, the necessary maps are
non-local \cite{Joness}. Moreover, by our construction,
the reality of the spectrum is robust insofar as its ${\cal CPT}-$symmetry
cannot be spontaneously broken. In this sense, our
example (\ref{63}) may be perceived as a
${\cal PT}-${\em asymmetric}  parallel to the
${\cal PT}-$symmetric quartic oscillator of Buslaev and Grecchi
\cite{BG}.

\subsection{The problem of boundedness}

Let us now turn to the operator $\mathcal F=C\mathcal P$. In order
to prove the invertibility of $\mathcal F$ and the boundedness of
${\mathcal {F}}^{-1}$ on $L^2 (\mathbf R)$ it is enough to
demonstrate the same facts for $C$. Factorization
(\ref{factorization}) implies that it will suffice to prove that
$C_1 =(\frac{d}{dx}+\frac{G}{2})$ is invertible and that
$C_1^{-1}$ is bounded on $L^2 (\mathbf R)$ if $G$ is given by
(\ref{eq6}). Indeed, we have
 \be C_1 =\frac{d}{dx}+\frac{1}{2}ax^2 +\frac{i}{2}bx
 \ee
and we now proceed as in Ref.~\cite{Ca04}. More precisely
 \be C_1 =\frac{d}{dx}+\frac{a}{2}\left(x+\frac{ib}{2a}\right)^2
  +\frac{b^2}{8a}
 \ee
is similar to
 \be C_2 =\frac{d}{dx}+\frac{a}{2}x^2 +\frac{b^2}{8a}
 \ee
via the complex translation $x\to x-\frac{ib}{2a}$. Hence $C_1$
has the same spectrum as $C_2$. In turn $C_2$ is unitarily
equivalent, via the Fourier transformation, to
 \be C_3 =-\frac{a}{2}\frac{d^2}{dx^2}+ix+\frac{b^2}{8a}.
 \ee
Therefore $C_1$ has the same spectrum as $C_3$. Finally, we
perform the unitary dilation $(U \psi)(x)=({a}/{2})^{1/6}\psi
[({a}/{2})^{1/3} x]$ and obtain that $C_1$ has the same spectrum
as
 \be
  C_4 =U C_3 U^{-1}=\left(\frac{a}{2}\right)^{1/3}
  \left[-\frac{d^2}{dx^2}+ix+
\left(\frac{a}{2}\right)^{-1/3}\frac{b^2}{8a}\right].
 \ee
Now, since the Schr\"odinger operator $-\frac{d^2}{dx^2}+ix$ has
an empty spectrum (see Ref.~\cite{He79}), so does $C_1$. In
particular $z=0$ belongs to resolvent set of $C_1$, so that $C_1$
is invertible and its inverse is bounded and defined on the whole
of $L^2 (\mathbf R)$.

\section{Towards operators ${\cal
C}$ of any finite order \label{ego}}

%\section{ Polynomial Charge Operators and Hermitian
% ``Metric'' Operators}
\label{Charge}

We shall postulate that the charge-operator component ${\cal C}$
of the pseudo-metric $ {\cal CP}$, where ${\cal P}$ denotes
parity, is a polynomial of any finite degree $\omega=0, 1, \ldots$
in the momentum operator $p$,
  \be
 {\cal C} = \sum_{k=0}^{\omega}\,
 \gamma_k(x)\,\frac{d^k}{dx^k},
 \ \ \ \ \ \ \ \ \gamma_k (x)= \gamma^R_k (x) + i\,\gamma^I_k (x) \,.
 \label{ansatz}
  \ee
The functions $ \gamma^R_k (x)$ and $\gamma^I_k (x) $ are both
assumed real, and our main task here is just to guarantee, at any
integer $\omega$, that the operator candidate for the metric $
{\cal CP}$ is Hermitian.

\subsection{The metric ${\cal CP}$ in differential form}

From
  \be
 {\cal C}^\dagger
 = \sum_{k=0}^{\omega}\,(-1)^k\, \sum_{\ell=0}^{k}\,
 \left (
 \ba
  k \\
 \ell
 \ea
 \right )
 \left [ \frac{d^{(k-\ell)}}{dx^{(k-\ell)}}\,
 \gamma^*_k(x)
 \right ] \, \frac{d^\ell}{dx^\ell}\, =
 \label{uansatz}
  \ee
  \ben
 = \sum_{\ell=0}^{\omega}\,(-1)^{\ell}\,
 \left \{
  \sum_{m=0}^{\omega-\ell}\, \,(-1)^{m}\,
 \left (
 \ba
  \ell+m \\
 \ell
 \ea
 \right )
 \left [
 \gamma_{\ell+m}^{R(m)}(x) - i\,
 \gamma_{\ell+m}^{I(m)}(x)
 \right ] \, \right \}\,\frac{d^\ell}{dx^\ell}\,,
  \een
where the superscripts $(m)$ at the functions $\gamma^R$ and
$\gamma^I$ indicate their $m-$tuple differentiation, one obtains
that the Hermiticity condition ${\cal CP}={\cal PC}^\dagger$ is
equivalent to the $(\omega+1)-$plet of relations
  \ben
 {\cal P}\,\gamma_\ell\,{\cal P} =\gamma^R_\ell (-x)
  + i\,\gamma^I_\ell
 (-x)=
 \sum_{m=0}^{\omega-\ell}\, \,(-1)^{m}\,
 \left (
 \ba
  \ell+m \\
 \ell
 \ea
 \right )
 \left [
 \gamma_{\ell+m}^{R(m)}(x) - i\,
 \gamma_{\ell+m}^{I(m)}(x)
 \right ] \,
  \een
with a trivial decoupling into its real and imaginary parts
  \be
 \gamma^R_\ell (-x)-\gamma^R_\ell (+x) =
 \sum_{m=1}^{\omega-\ell}\, \,(-1)^{m}\,
 \left (
 \ba
  \ell+m \\
 \ell
 \ea
 \right )
 \gamma_{\ell+m}^{R(m)}(x)
 \,
 \label{Reca}
  \ee
and
  \be
 \gamma^I_\ell (-x)+ \gamma^I_\ell (+x)=-
 \sum_{m=1}^{\omega-\ell}\, \,(-1)^{m}\,
 \left (
 \ba
  \ell+m \\
 \ell
 \ea
 \right )
  \gamma_{\ell+m}^{I(m)}(x)
  \,,
 \label{recb}
  \ee
respectively, with $\ell = \omega-k=0, 1, \ldots, \omega$.

\subsection{Functional freedom in
 complex coefficients $\gamma_k(x)$}

At the first few $k=0, 1, \ldots$ the above Hermiticity
constraints degenerate to the comparatively elementary relations,
  \ben
 \gamma^R_\omega (x)-\gamma^R_\omega (-x) = 0, \ \ \ k = 0,
  \een
  \ben
 \gamma^R_{\omega-1} (x)-\gamma^R_{\omega-1} (-x) =
 \left (
 \ba
  \omega \\
 1
 \ea
 \right )
 \gamma_{\omega}^{R(1)}(x)
 \,, \ \ \ \ \ k = 1,
  \een
  \ben
 \gamma^R_{\omega-2} (x)-\gamma^R_{\omega-2} (-x) =
 \left (
 \ba
  \omega-1 \\
 1
 \ea
 \right )
 \gamma_{\omega-1}^{R(1)}(x)
 -
 \left (
 \ba
  \omega \\
 2
 \ea
 \right )
 \gamma_{\omega}^{R(2)}(x)
 \,, \ \ \ k = 2,
  \een
etc, or, in parallel,
  \ben
 \gamma^I_{\omega} (x)+ \gamma^I_{\omega} (-x)=0, \ \ \ k = 0,
  \een
  \ben
 \gamma^I_{\omega-1} (x)+\gamma^I_{\omega-1} (-x) =
 \left (
 \ba
  \omega \\
 1
 \ea
 \right )
 \gamma_{\omega}^{I(1)}(x)
 \,, \ \ \ \ \ k = 1,
  \een
  \ben
 \gamma^I_{\omega-2} (x)+\gamma^I_{\omega-2} (-x) =
 \left (
 \ba
  \omega-1 \\
 1
 \ea
 \right )
 \gamma_{\omega-1}^{I(1)}(x)
 -
 \left (
 \ba
  \omega \\
 2
 \ea
 \right )
 \gamma_{\omega}^{I(2)}(x)
 \,, \ \ \ k = 2,
  \een
etc. This means that the symmetric parts $H_\ell (x)=H_\ell (-x)$
of all $\gamma^R_\ell (x)$ are arbitrary functions while, in
parallel, the antisymmetric parts $h_\ell (x)= -h_\ell (-x) $ of
all $\gamma^I_\ell (x)$ are also arbitrary. We may conjecture that
the remaining components $R_\ell (x)=\gamma^R_\ell (x)-H_\ell
(x)=-R_\ell (-x)$ and $r_\ell (x)=\gamma^I_\ell (x)-h_\ell
(x)=r_\ell (-x)$ obey the rules
  \be
 R_\omega=0, \ \ \  R_{\omega-1} (x) =
 \frac{ \omega }{2}
 H_{\omega}^{(1)}(x), \ \ \ R_{\omega-2} (x) =
 \frac{ \omega-1 }{2}
 H_{\omega-1}^{(1)}(x), \ \ \ \ldots
 \label{former}
  \ee
while
  \be
 r_\omega=0, \ \ \  r_{\omega-1} (x) =
 \frac{ \omega }{2}
 h_{\omega}^{(1)}(x), \ \ \ r_{\omega-2} (x) =
 \frac{ \omega-1 }{2}
 h_{\omega-1}^{(1)}(x), \ \ \ \ldots\ .
  \ee
and are fully determined by the respective recurrent relations
(\ref{Reca}) and (\ref{recb}).

\subsection{Proof}

We see that both the sequences $R_{\omega-k} (x)$ and
$r_{\omega-k} (x)$ have precisely the same structure so that just
the sequence of $R_{\omega-k} (x)$ may be considered without any
loss of generality. Its elements should be evaluated in the
recurrent manner with respect to the growing $k$. The appropriate
\textit{Ans\"atze} may be written in the finite-series form where,
formally, $ H_{\omega+1} (x)= H_{\omega+2} (x)= \cdots = 0$ and $
h_{\omega+1} (x)=  h_{\omega+2} (x)= \cdots = 0$,
  \be
 \gamma^R_{\omega-k} (x) =
  H_{\omega-k} (x)
 + \sum_{m=1}^k c_m\,\frac{(\omega-k+m)!}{(\omega-k)!}
    H_{\omega-k+m}^{(m)} (x)\,,
  \ee
  \be
 \gamma^I_{\omega-k} (x) =
  h_{\omega-k} (x)
 + \sum_{m=1}^k c_m\,\frac{(\omega-k+m)!}{(\omega-k)!}
    h_{\omega-k+m}^{(m)} (x)\,.
  \ee
With an auxiliary $c_0=1$ these \textit{Ans\"atze} describe all
the $\omega-$dependence of our functions $\gamma
=\gamma^R+i\gamma^I$ in closed form.

As already stated above, the first term and the subsequent sum are
of an opposite parity in both these formulae since $c_{2n}=0$ at
all $n = 1, 2, \ldots$. This observation is easily proved since
after the insertion of the latter two \textit{Ans\"atze},  the
complicated recurrences (\ref{Reca}) are replaced by their
simplified version
  \ben
 2c_1=\frac{c_0}{1!}, \ \ \ 2c_2=\frac{c_1}{1!}-\frac{c_0}{2!}, \ \
 \ \ldots\,,
  \een
\textit{i .e.},
  \be
 2c_k=\sum_{m=0}^{k-1} \,(-1)^{k-m-1}\,\frac{c_m}{(k-m)!}\,.
 \label{reur}
  \ee
It is worthwhile to point out that the $c_k$ coefficients with odd
$k$ can be written in terms of Bernoulli numbers (see, {\it e.g.},
Ref.~\cite{AS72})
 \be
 c_{2n-1}=\frac{2(2^{2n}-1)}{(2n)!}B_{2n}\quad (n>0)\,.
 \ee
The key idea of an explicit solution of these recurrences is that
the generating function $f(x)=\sum \,c_kx^k$ of the coefficients
$c_m$ must satisfy the functional equation $f(x) - 2 = -f(x) /e^x$
which is, in its turn, easily solvable. In this way we arrive at
the solution of recurrences (\ref{reur}) in the following compact
form,
  \be
 f(x) = c_0+c_1x+c_2x^2+\ldots = \frac{2}{1+\exp (- x)}= 1+ \tanh
 \frac{x}{2}=
  \ee
  \ben
   = 1+\sum_{n=1}^{\infty}\frac{2(2^{2n}-1)}{(2n)!}B_{2n}
   x^{2n-1}\,.
  \een
Obviously, all the possible parity-violating terms in the
right-hand side of our Hermiticity conditions (\ref{Reca}) vanish.
This makes the form of our polynomial charge ${\cal C}$ extremely
flexible and confirms the consistency of its present construction.

We may conclude that the requirement of Hermiticity of the metric
${\cal CP}$ defines all the antisymmetric components $R_\ell(x)$
and their spatially symmetric partners $r_\ell(x)$. It does not
impose any additional constraint either upon the real and
spatially symmetric coefficient functions $H_\ell(x)$ or upon
their purely imaginary spatially antisymmetric partners
$h_\ell(x)$.

\section{Outlook \label{five}}

We now sketch some possible applications of our methods to a
variety of problems where quasi-Hermitian or pseudo-Hermitian
operators are involved.

In the context of the Klein-Gordon description of the free motion
of a spinless particle in the ``usual" Hilbert space ${\cal H}$
the relativistic evolution is generated by the Feshbach-Villars
\cite{FV} ``Hamiltonian" $H_{(FV)}$ which proves non-Hermitian,
 \be
 |\psi(t)\rangle = e^{-iH_{(FV)}(t-t_0)}|\psi(t_0)\rangle
 , \ \ \ \ \ \ \
 H_{(FV)}=
 -\frac{1}{2}
 \left (
 \begin{array}{cc}
 1-\triangle&-\triangle\\
 \triangle&\triangle-1
 \ea
 \right )
 \label{TE}
 \ee
(cf., e.g., p. 341 in ref. \cite{Constantinescu}). One should
notice that this model works with the differential
pseudo-Hermitian operator with structure which strongly resembles
the usual Schr\"{o}dinger operators in the simplest non-trivial,
two-dimensional coupled-channel case. Thus, we may expect that the
methods described in our previous study might find an immediate
extension to the similar problems.

The idea may also find applications in a broader context, say, of
the boson mappings in nuclear physics which were comprehensively
discussed in the paper \cite{Geyer}. It is shown there that a
consistent quantum mechanical framework, and in particular a
viable variational calculation for non-hermitian Hamiltonians, can
indeed be constructed after the introduction of a non-trivial
metric. In the context of Holstein-Primakoff type mappings this
freedom defines the link with so-called Dyson-Maleev type mappings
(see \cite{dobaczew}). In practical computations, a puzzling
non-Hermiticity of observables proved more then compensated by the
advantages, as has been amply demonstrated in applications of
generalized Dyson-Maleev mappings (see \cite {gey2004} and
references cited therein).

All the technical conditions imposed upon the ``true physical
metric" $\Theta$ in review \cite{Geyer} are important, especially
if one tries to work within a truly infinite-dimensional Hilbert
space. This has been emphasized by Kretschmer and Szymanowski
\cite{Kr04} who showed that the use of the toy metric operators
might require a careful scrutiny because these operators remain
unbounded. In this context, ref. \cite{Ca04} as well as our
present paper demonstrated persuasively that a switch to the use
of the differential operators ${\cal C}$ might be understood as an
important new idea.

All the similar observations must be perceived as individual steps
of a systematic improvement of the mathematically correct
understanding of the use of the differential operators in
connections with many applications of the quasi-Hermitian
observables which seems to range, at present, from the elementary
descriptions of the localization transitions in solid state
physics \cite{Hatano} up to many ambitious ${\cal PT}-$symmetric
models in quantum field theory \cite{Jcz}.

The experience gained during our study of the simple
Schr\"{o}dinger equations might equally well find applications on
the very boundary of quantum mechanics (like, say, in cosmology
\cite{Alki}) or even in the domain of the classical model-building
(e.g., in the magneto-dynamics of fluids \cite{Uwe}) and in the
various physical models of different origin characterized by the
simple matrix structure of their description (see a number of
their most elementary samples mentioned in the short and nice
review \cite{Berry}) where the eigenvalues coalesce or almost
coalesce in the manner which contradicts the standard and robust
finite-dimensional Hermitian-matrix mathematics. Of course, all
these mathematical problems and not entirely standard physical
situations may impose new and challenging tasks and motivate a
deeper future analysis of the questions outlined in our present
paper.

\newpage

\section*{Acknowledgements}

We are grateful to our colleague A. Ventura who actively
participated in the initial stage of this work. A. B. wishes to
thank the University Grants Commission,New Delhi, for the award of
a Junior Research Fellowship. C. Q. is a Research Director,
National Fund for Scientific Research (FNRS), Belgium. M. Z.
(supported by the grant Nr. A 1048302 of GA AS CR) appreciates the
inspiring atmosphere and hospitality of INFN and Dipartimento di
Fisica, Universit\`{a} di Bologna and of the Physics Department of
the University of Stellenbosch.

\end{document}